\pdfoutput=1
\documentclass[twocolumn,english,aps,superscriptaddress,prb,nofootinbib]{revtex4-2}

\usepackage{babel}
\usepackage{amsmath}
\usepackage{amssymb}
\usepackage{graphicx}
\usepackage{xcolor}
\usepackage{tikz}
\usepackage{hyperref}

\begin{document}

\title{Weakly first-order melting of the 1/3 plateau in the Shastry-Sutherland model}

\author{Samuel Nyckees}
\affiliation{Institute of Physics, Ecole Polytechnique F\'ed\'erale de Lausanne (EPFL), CH-1015 Lausanne, Switzerland}
\author{Philippe Corboz}
\affiliation{Institute for Theoretical Physics and Delta Institute for Theoretical Physics, University of Amsterdam, Science Park 904, 1098 XH Amsterdam, The Netherlands}
\author{Fr\'ed\'eric Mila}
\affiliation{Institute of Physics, Ecole Polytechnique F\'ed\'erale de Lausanne (EPFL), CH-1015 Lausanne, Switzerland}

\date{\today}
\begin{abstract} 

We investigate the thermal properties of the 1/3 plateau in the Shastry-Sutherland model with infinite projected entangled-pair states (iPEPS) by performing the imaginary time evolution of the infinite temperature density matrix. We show that both the $\mathbb{Z}_2$ and $\mathbb{Z}_3$ broken symmetries of the ground states are restored at a unique temperature where the correlation length has a peak, and that the melting of the plateau occurs via a single weakly first-order transition. We focus on the experimentally relevant coupling constants deep into the 1/3 plateau phase at $h = 1$ and $J'/J = 0.63$, which was estimated to describe the SrCu$_2$(BO$_3$)$_2$ compound. By computing the free energy we are able to locate the transition temperature around $T_c \simeq 4.8$K well above the temperature $T=2$K, at which the 1/3 plateau was observed in experiments. The investigation is supplemented by adding a bias term to the Hamiltonian and studying the induced crossover. We further map the transition line in the field-temperature phase diagram.

\end{abstract}

\maketitle


\section{Introduction}

\par
\begin{figure}[t!]
\includegraphics[width=0.40\textwidth]{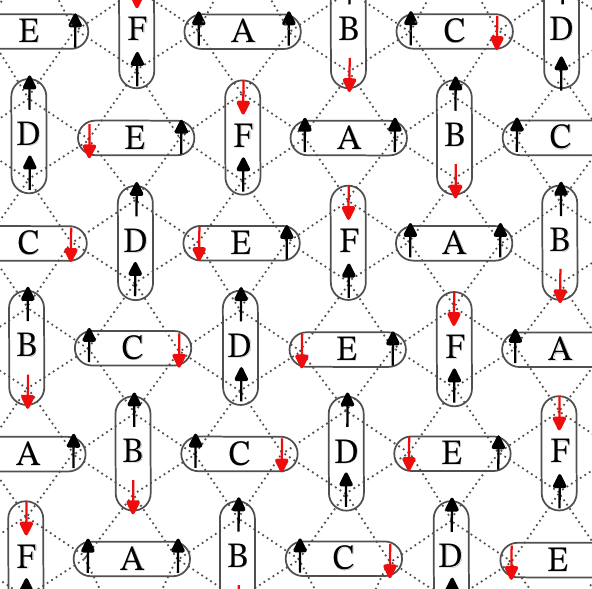}
\caption{The Shastry-Sutherland lattice where the dotted lines represent the inter-dimer Heisenberg interactions with coupling constant $J'$. The intra-dimer Heisenberg interaction $J$ (not shown here) takes place between two spins of the same dimer. The black oval shapes represent the dimers. The ground state of the 1/3 plateau has a six-sublattice dimer structure, here denoted by $A,B,C,D,E$, and $F$. We represent one of the six degenerate ground states where the triplet up-up state lies on the sub-lattices $A$ and $D$, the down-up or up-down states on the other sub-lattices. The other five ground states are obtained by translation and rotation by $\pi/2$.}
\label{fig:lattice}
\end{figure}

 Since the discovery of its experimental realisation in the SrCu$_2$(BO$_3$)$_2$ compound\cite{Ueda1999a,Ueda1999b,Ueda2003}, the Shastry-Sutherland model (SSM) \cite{shastry1981} has become a rich playground for both condensed matter theorists and experimentalists. The Hamiltonian of the SSM is given by: 
\begin{align}
    H_0 =  J' \sum_{\langle\langle i,j \rangle\rangle} \vec{S}_i \cdot \vec{S}_j+ J \sum_{\langle i,j \rangle} \vec{S}_i \cdot \vec{S}_j - h \sum_i S_i^z
\end{align}
\noindent where $J'$ and $J$ denote the coupling constants and play the role of inter-dimer and intra-dimer interactions, respectively, as shown in Fig. \ref{fig:lattice}. Moreover, $\vec{S}_i$ are spin-1/2 operators and $h$ denotes the external magnetic field. The position of the Cu $^{2+}$ ions in the crystal structure of the SrCu$_2$(BO$_3$)$_2$ reproduces the Shastry-Sutherland lattice and interact via exchange couplings $(J, J')$, the ratio of which at ambient pressure, is estimated to be around $J'/J \simeq 0.63$ \cite{Mila2013b}.

\par In the absence of an external magnetic field, it is well established that the model has an exact dimer phase in the small $J'/J$ limit and a Néel ordered phase in the large $J'/J$ limit, with a plaquette phase in between ~\cite{Kawakami2000,Nakano2018,Ashvin2019,Mila2013a}. Some recent studies ~\cite{Ling2022,Wang_2022,Erhai2022,Fa2024} further suggested the existence of a spin liquid phase separating the plaquette and the Néel ordered phases. 

\par Upon the introduction of an external magnetic field, the SrCu$_2$(BO$_3$)$_2$ compound exhibits a series of magnetization plateaus at low temperatures \cite{Mila2004,Levy_2008,Gaulin2012,Rosenbaum2016,Haravifard2019}, which have also been found numerically in the SSM at zero temperature \cite{Goto2000,Gaulin2004,Bruce2012,Mila2013b,Mila2013c,Gaulin2008}. Among these, the two largest are the 1/2 and 1/3 plateaus, whose structures are relatively simple and well understood. The 1/2 plateau structure of is made of a checkerboard of triplet dimers and has been shown to melt via an Ising transition \cite{Jacek2021}. The 1/3 plateau spin structure is made of alternating up-up, up-down and down-up dimers as shown in Fig. \ref{fig:lattice}. The ground state is six fold degenerate and breaks the $\mathbb{Z}_3$ translational and $\mathbb{Z}_2$ rotational symmetries. 

\par The 1/3 plateau is thus expected to melt either in a two step process where the $\mathbb{Z}_3$ and $\mathbb{Z}_2$ symmetries are restored at different temperatures or via a single transition where the $\mathbb{Z}_2$ and $\mathbb{Z}_3$ symmetries are both restored at the same temperature. If both symmetries are restored simultaneously, the model is expected to follow a six-state Potts-like scenario where the melting occurs through a weakly first-order transition. It is worth noting that the classical Ising model on the Shastry-Sutherland lattice exhibits a 1/3 plateau as well, whose ground state has the same degeneracy and breaks the same symmetries \cite{Dublenych2012,Fong2009}. 

\par In this work we perform a thorough investigation of the melting of the 1/3 plateau using infinite projected entangled-pairs state (iPEPS) methods \cite{Cirac2008, Gang2008, Xiang2008, Nishio2004}. These methods rely on expressing the thermal density matrix (or the wave function) as a 2D tensor network whose control parameter is the dimension of the local tensors, usually referred to as the bond dimension and denoted by $D$. Two different approaches are taken. First, the free energy is computed for different bond dimensions and extrapolated with respect to the truncation error. In the second approach, a bias term is introduced to the Hamiltonian and the scalings of the order parameter and its temperature derivative are analysed. Both approaches lead to a consistent picture where the melting occurs via a weakly first-order transition around $T_c \simeq 0.059$. 

\par The paper is organized as follows. In Section II we recall the thermal iPEPS ansatz and introduce a way to compute the free energy. In Section III we discuss the approach based on adding a pinning field to the Hamiltonian which biases the model and favours one of the ground states. In Section IV we discuss various results obtained with and without the pinning field and map the melting of the 1/3 plateau in the finite temperature phase diagram. Finally, in Section V we summarize our main conclusion. Moreover, we provide a brief discussion of the classical model in Appendix B, where we also report the melting to occur via a weakly first-order transition. 

\section{Numerical method}

The success of the matrix product state (MPS) formalism \cite{Schollwock2011} led to its two-dimensional generalisation and in particular to the development of the projected entangled-pairs state (PEPS) \cite{Cirac2004a, Cirac2004b}, and later on, its infinite counterpart (iPEPS) \cite{Cirac2008, Gang2008, Xiang2008, Nishio2004}. Initially introduced as an ansatz for two-dimensional ground state wavefunctions it can also be used to simulate thermal Gibbs ensembles \cite{Corboz2019,Jacek2018,Li2011,Czarnik2012,Kshetrimayum2019,Jimenez2021}. We now briefly recall the iPEPS ansatz formalism and how to compute observables such as the free energy.

\subsection{Gibbs ensemble with iPEPS}

\par
\begin{figure}[t!]
\includegraphics[width=0.45\textwidth]{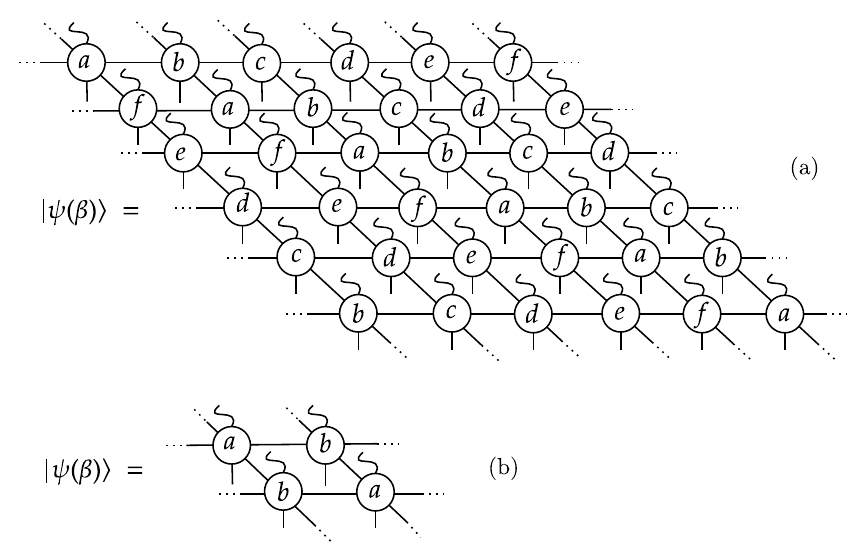}
\caption{We show two different iPEPS ansatz for the purified state with different unit cells. We refer to (a) as the $6\times 6$ unit cell and to (b) as the $2\times 2$ or bi-partite unit cell. The local tensors ($a,b,c,d,e,f$) have dimensions $D \times D \times D \times D \times 4 \times 4$ where the wiggled legs and the vertical legs represent the ancilla and the physical space with dimensions $4$, respectively. The legs connecting the local tensors with their neighboring tensors represent the virtual legs with dimension $D$.}
\label{fig:peps_unit_cell}
\end{figure}

 At zero temperature, the iPEPS ansatz represents 2D ground state wavefunctions as a tensor network on the square lattice with arbitrary unit cell of size $N_x \times N_y$ composed of local tensors $A^{[x,y]}_{\alpha, \beta, \delta, \gamma, i }$, where $(x,y)$ denotes the position of the tensor within the unit cell. The index $i$ is referred to as the physical leg and has the same dimension as of the local Hilbert space while the indices $\alpha, \beta, \delta, \gamma$ are referred to as the virtual legs and have dimension $D$ called the bond dimension. It is possible to write the Gibbs ensemble using the same formalism by expressing the thermal density matrix as a purified state such that 
\begin{align}
\rho(\beta) & \equiv e^{-\beta H} \nonumber \\
& = \text{Tr}_s (| \psi (\beta) \rangle \langle  \psi (\beta) |
\end{align}
where the purified state $| \psi (\beta) \rangle$ is written as a tensor network on the square lattice with tensors $A^{[x,y]}_{\alpha, \beta, \delta, \gamma, i ,s}$. The tensors used at finite temperature have one more index $s$, usually referred to as the ancilla leg, which acts on the purifying Hilbert space. Similarly to Ref. \cite{Jacek2021}, we choose to work with the dimer setup, where one tensor per dimer is used instead of one tensor per spin. We illustrate in Fig. \ref{fig:peps_unit_cell} two representations for $|\psi(\beta) \rangle$ with different unit cells relevant for this work.

Provided $ \text{Tr}_s ( | \psi (0) \rangle \langle  \psi (0) |) = \mathbb{I}$, the thermal ensemble can then be written as an imaginary time evolution of the infinite temperature purified state: 
\begin{align}
\rho(\beta) & = \text{Tr}_s ( e^{-\beta H /2 }| \psi (0) \rangle \langle  \psi (0) | e^{-\beta H/2} )  
\end{align}
and the initial state at infinite temperature can be written as
\begin{align}
| \psi (0) \rangle = \prod_{i \in \text{dimer}}\sum_{s_i =1}^{4} | s_i, s_i \rangle.
\end{align}

\par To perform the imaginary time evolution of the purified state we use the simple update (SU) scheme \cite{Xiang2008} which we now briefly recall. The evolution of the purified state is performed by applying the Trotter-Suzuki decomposition of $e^{-\beta H}$ onto the state $|\psi(0)\rangle$. During the initial time steps, each bond dimension of the local tensors increases until it reaches the threshold value $D$ and the application of each gate is computed exactly. Past that point, after the application of each gate, the local tensor's bond dimensions are truncated down to the bond dimension $D$ by performing a singular value decomposition and keeping only the $D$ largest singular values. A more detailed explanation of the simple update scheme can be found in Ref. \cite{Xiang2008}. There exist more sophisticated schemes to perform the imaginary time evolution, such as the neighbourhood tensor update \cite{Jacek2021b} or the full / fast-full update (FU /FFU) \cite{jordan2008,Orus2015} which are more accurate but also computationally more expensive. We discuss and compare the FFU with the SU in Appendix \ref{section_ffu}.

\par Local observables $\langle O(\beta) \rangle$ are computed by approximately contracting the infinite 2D tensor network. This task is performed by the Corner Transfer Matrix Renormalisation Group (CTMRG) \cite{Baxter1968,Okunishi1996,Vidal2009,Troyer2011,Troyer2014} algorithm which approximates the infinite tensor network by different environment tensors $E_{[x,y]} = \{C_{[x,y]}^{\alpha}, T_{[x,y]}^{\alpha} | \alpha \in \{1,2,3,4\} \}$ where $i$ denotes one of the local tensors in the unit cell. The environments are composed of row $T_{[x,y]}^\alpha$ and corner $C_{[x,y]}^\alpha$ tensors of dimension $\chi \times D^2 \times \chi$ and $\chi \times \chi$ respectively where $\chi$, which from now on will be referred to as the boundary bond dimension, acts as the control parameter, as in the infinite $\chi$ limit the algorithm recovers the exact contraction. One major advantage of the algorithm is that it gives direct access to the transfer matrix and hence to the correlation length. We refer the reader to Ref. \cite{Troyer2014} for more details on the algorithm. 

\subsection{Free energy}

\par
\begin{figure}[t!]
\includegraphics[width=0.45\textwidth]{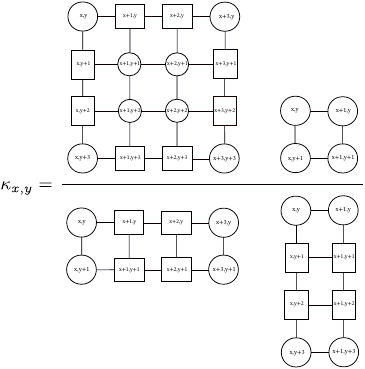}
\caption{Partition function per site computed with the CTMRG algorithm. The circles and rectangles represent the square transfer matrices and row/columns tensors respectively.}
\label{fig:free_ener_ctm}
\end{figure}

 Although its modern formulation was introduced by Okunishi and Nishino, the first application of the CTMRG algorithm can be traced back to Baxter \cite{Baxter1968,Baxter1978} in the context of 2D statistical physics where the partition function can be written as the contraction of an infinite 2D tensor network. Baxter further derived the one-site contribution to the partition function $\kappa = Z^{1/N}$  for a uniform tensor network. This result can be extended to the contribution of one unit cell to the partition function by introducing $\kappa_{x,y}$, illustrated with tensor network notation in Fig. \ref{fig:free_ener_ctm}, and setting:
\begin{align}
\kappa^4 = \prod_{x,y \in \text{unit cell}} \kappa_{x,y}.
\label{kappa}
\end{align}
The power 4 accounts for the fact that $\kappa_{x,y}$ is the contribution to the partition function of a unit cell of size $2\times2$ as shown in Fig. \ref{fig:free_ener_ctm}. For thermal iPEPS, the free energy per dimer is given by:
\begin{align}
f = -\frac{T}{N_{\text{dimer}}} \log (\text{Tr} (\rho(\beta)) ) 
\end{align}
where $\text{Tr} (\rho(\beta))$ is written as the contraction of an infinite two-dimensional tensor network and the free energy can be computed using Eq. \ref{kappa} with:
\begin{align}
f = -T\log(\kappa).
\label{kappa2}
\end{align}
In this work, we focus on two different unit cells shown in Fig. \ref{fig:peps_unit_cell}. The free energy per dimer using the $6 \times 6$ unit cell is computed with $\log(\kappa) = (\log(\kappa_{1,1}) + \log(\kappa_{1,3}) + \log(\kappa_{1,5}))/12)$ while the free energy per dimer using the $2\times 2$ unit cell is computed with $\log(\kappa) = \log(\kappa_{1,2})/4$. It is worth noting that, in order to recover the density matrix with the right pre-factor, one has to take into account the renormalisation factor applied during the time evolution.

\section{pinning field}
\par In order to investigate the nature of the melting, we can study the crossover induced by adding an extra pinning field $h_s>0$ which favours one of the ground states. By using the structure of the ground state and dividing the square lattice into six different sub-lattices denoted by $A,B,C,D,E$ and $F$ as shown in Fig. \ref{fig:lattice}, we introduce the pinning field as follows: 
\begin{align}
    H =  H_0 - h_s \sum_{i \in A,D} \tilde S_i^z + \frac{h_s}{2} \sum_{i \in B,E,C,F}  \tilde S_i^{z}
\end{align}
where $\tilde S_i^z$ denotes the total magnetization on the dimer at position $i$. By denoting $m_\alpha$, the magnetization on the sub-lattice $\alpha$, we can derive the order parameter associated to the pinning field $h_s$, and by using the symmetry of the model in the 1/3 plateau, $m_A = m_D$ and $m_B = m_E = m_C = m_F$, we obtain:
\begin{align}
 m &  = m_A + m_D + e^{2 i\pi /3 }(m_B + m_E) \nonumber \\ 
& + e^{-2 i\pi /3 }(m_C + m_F).
\end{align}
We recover the order parameter associated to the breaking of the $\mathbb{Z}_3$ translational symmetry and we define $\psi_{Z_3} = m/6$.  

\par For continuous transitions, the order parameter scales like:
\begin{align}
\psi_{Z_3}(t,h_s) = h_s^{(d-y_h)/y_h} f(h_s^{-y_t/y_h}t) \label{scaling0}
\end{align}
with $t$ the reduced critical temperature and $y_t$ and $y_h$ the thermal exponents which are related to the usual critical exponents that characterise the divergence of the correlation length $\nu$ and decay of the order parameter $\beta$, via $y_t = 1/\nu$ and $d-y_h = \beta / \nu$. $f$ is a scaling function. For first-order transitions in two dimensions, it has been shown that the scaling still holds but with thermal exponents $y_h = 2$ and $y_t = 2$ or equivalently critical exponents $\beta = 0$ and $\nu = 1/2$ \cite{Landau1984,Fisher1982} and the scaling (\ref{scaling0}) becomes:
\begin{align}
& \psi_{Z_3}(t, h_s) =  f(h_s^{-1}t). \label{scaling}
\end{align}
Hence, upon the introduction of the pinning field, the order parameter, which jumps at the critical temperature at zero field, will smoothen out with a maximum slope proportional to $h_s^{-1}$. The temperature at which the order parameter changes convexity is called the crossover temperature and is denoted by $T^*(h_s)$. The scaling further implies that the crossover temperature approches the critical temperature like:
\begin{align}
    T^*(h_s) - T_c \propto h_s^{-1}.
\end{align}

\section{Results}

\par We now discuss the results obtained with and without the pinning field. For all simulations, we consider the experimental relevant ratio at $J'/J = 0.63$ and $J = 1$. For most of the paper we will focus on the particular cut of the finite temperature phase diagram at $h = 1$ which is located in the middle of the 1/3 plateau \cite{Mila2013b}. All simulations were performed with a time step of $\Delta \beta = 10^{-3}$. We used the ITensor \cite{itensor1} library and in particular its U(1) symmetry implementation to improve the efficiency of the simulations.

\subsection{Without pinning field}
\label{bc_sec}

\par
\begin{figure}[t!]
\includegraphics[width=0.45\textwidth]{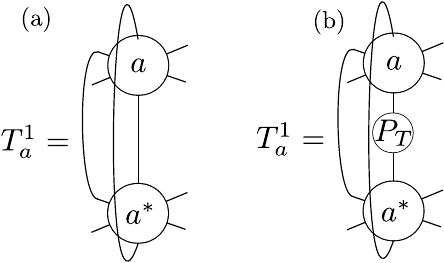}
\caption{Different initialisations of the row tensor $T^1_a$ based on the bulk tensor $a$. (a) Standard initialisation of the row tensor. (b) Fixed boundary condition where $T^1_a$ is initialized by projecting the local space of $a$ onto the up-up triplet. The generalisation to others tensors in the CTMRG environment is straightforward.}
\label{fig:bc_ctmrg}
\end{figure}

 In this section, as there is no pinning field, we choose to imaginary time evolve the purified state assuming the bipartite lattice. We then contract the tensor network with the CTMRG algorithm using the 2$\times$2 bipartite unit cell and the 6$\times$6 unit cell, and compare the results. If we contract exactly the density matrix with open boundary conditions, the order parameter should stay zero even within the ordered phase. However, due to various approximations in the contraction, the CTMRG algorithm usually breaks the symmetry and converges to one of the ordered environments. Here, instead of letting the CTMRG break the symmetry on its own, we fix the boundary conditions by projecting the initial boundary tensors on one of the broken symmetry ground states. In particular, we project the local tensors on the sub-lattices $A$ and $D$ on the triplet up-up state while the tensors on the sub-lattices $B,C,E$ and $F$ are projected onto the singlet state. We illustrate this in tensor notation in Fig. \ref{fig:bc_ctmrg} where the projectors are given by:
\begin{align}
P_T = \begin{pmatrix}
 	  1 & 0 & 0 & 0 \\
	  0 & 0 & 0 & 0 \\
	 0 & 0 & 0 & 0 \\
	 0 & 0 & 0 & 0 
\end{pmatrix},
\quad
P_S = \begin{pmatrix}
 	 0 & 0 & 0 & 0 \\
	 0 & \sqrt{2} & -\sqrt{2} & 0 \\
	 0 & -\sqrt{2} & \sqrt{2} & 0 \\
	 0 & 0 & 0 & 0 
\end{pmatrix}
\end{align}
if projected on the triplet up-up or on the up-down/ down-up subspace respectively. This in turn, improves greatly the convergence of the algorithm within the ordered phase. 

\par Moreover, the projection allows us to control the hysteresis effects in the vicinity of the first order transition. By using the $2 \times 2$ unit cell, the CTMRG algorithm enforces the state to be uniform while the projection in the $6 \times 6$ cell targets the ordered state. By comparing the free energy of the two states, we can accurately determine the location of the phase transition. Without the projection, $T_c$ may be underestimated due to the hysteresis behavior, since the state is evolved starting from the disordered phase at infinite temperature, and it may remain metastable even slightly below $T_c$.

\subsubsection{Order parameter}

\begin{figure}[t!]
\includegraphics[width=0.45\textwidth]{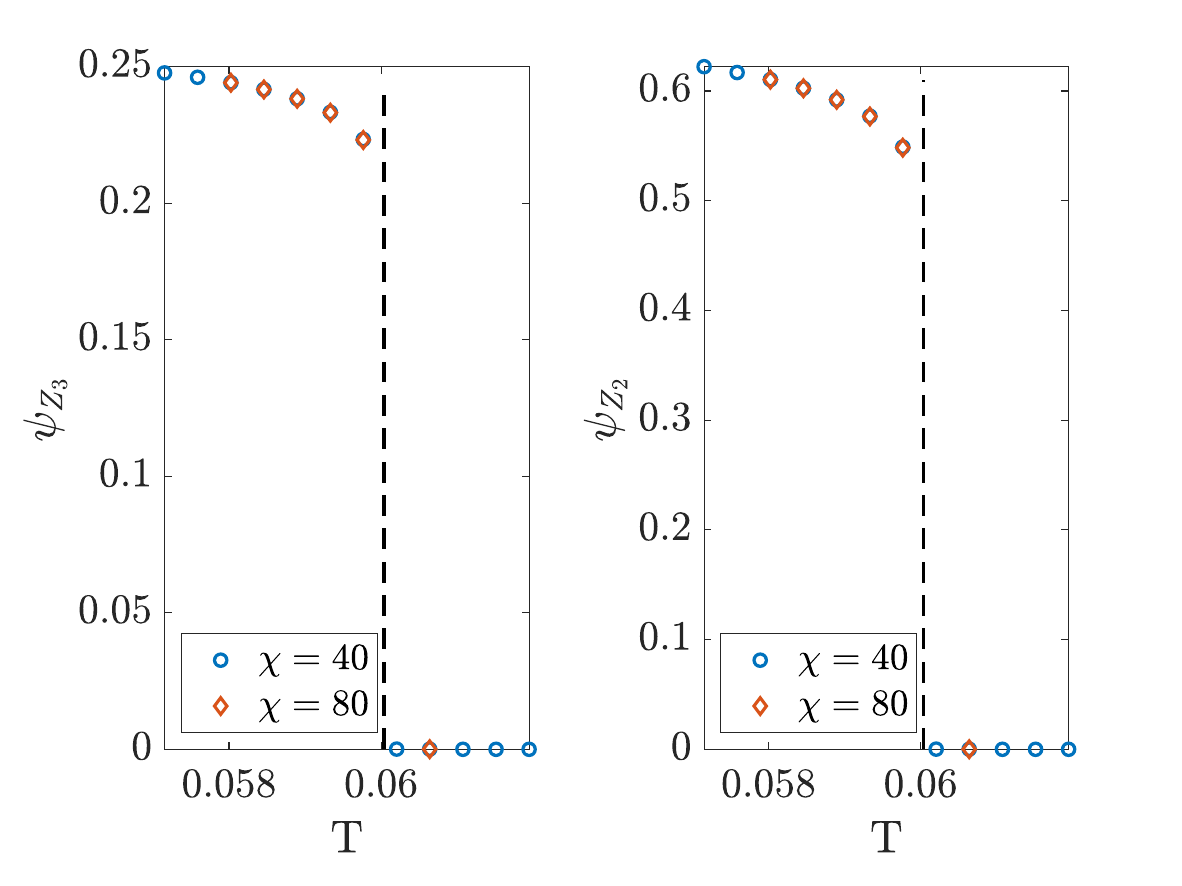}
\caption{$ \mathbb{Z}_3$ (left panel) and $ \mathbb{Z}_2$ (right panel) order parameters with respect to the temperature for different bond dimensions for $D = 13$. We can see that both symmetries are restored at a unique temperature in agreement with a first-order phase transition. The dashed line represents the critical temperature obtained from locating the kink in the free energy (see Fig. \ref{fig:D13_Tc}).}
\label{fig:order_paramD13}
\end{figure}

We introduce the $\mathbb{Z}_2$ order parameter as:
\begin{align}
\psi_{Z_2} = \langle \tilde S^{z}_{x,y} \tilde S^{z}_{x+1,y+1} \rangle -  \langle \tilde S^{z}_{x,y} \tilde S^{z}_{x+1,y-1} \rangle 
\end{align}
where the site $(x,y)$ lies on the sub-lattice $A$ and $\tilde S^z_{x,y}$ is again the total magnetization of the dimer on the site $(x,y)$. We show the results in Fig. \ref{fig:order_paramD13} for $D = 13$. We observe that both order parameters associated to the $ \mathbb{Z}_2$ rotational and $ \mathbb{Z}_3$ translational symmetry breaking are restored at a unique temperature. This is the first evidence for the ordered phase to melt via a direct first-order transition rather then a two-step process. We compare the results for different bond dimensions $\chi = 40$ and $\chi = 80$ to ensure that they have converged with respect to $\chi$.

\subsubsection{Free energy}

\par
\begin{figure}[t!]
\includegraphics[width=0.45\textwidth]{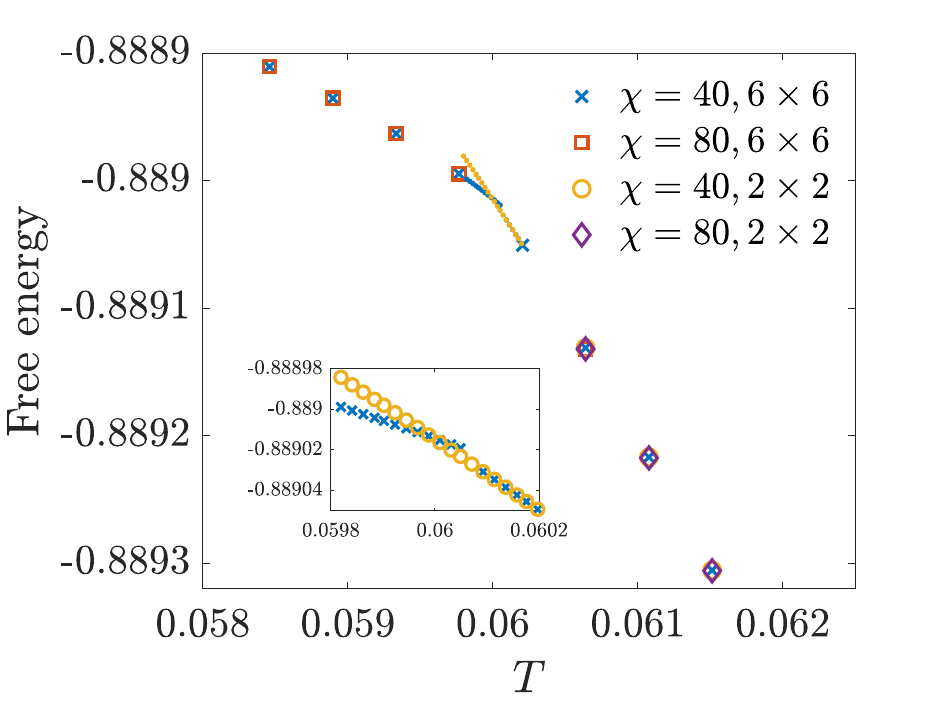}
\caption{Free energy for $D = 13$ for different bond dimensions $\chi$. The inset shows a zoom near the transition. We observe that far away from the transition in the disordered phase, using different unit cells lead to the same results while in the vicinity of the transition, using the $2\times 2$ unit cell imposes the state to be uniform and lead to hysteresis effect visible in the inset. The intersection of the free energies of the two states gives $T_c(D=13) = 0.060$.}
\label{fig:D13_Tc}
\end{figure}

 We now investigate the behavior of the free energy. The results obtained for $D = 13$ are shown in Fig. \ref{fig:D13_Tc} where we observe a distinct kink in the free energy around $T_c(D=13) \simeq 0.060$. This temperature agrees with the temperature at which both symmetries are restored, all in agreement with a first-order transition. We further check that the results have converged in $\chi$ by comparing different bond dimensions $\chi = 40$ and $\chi = 80$ and observing no difference. Far away from the transition in the disordered phase, the free energy is computed using the projected $6\times 6$ unit cell and using the $2\times 2$ unit cell give the same result. In contrast, near the transition, using the different unit cells leads to some hysteresis effect visible on the inset of Fig. \ref{fig:D13_Tc}. It is worth noting that in the disordered phase, using the $2\times 2$ unit cell converges much faster than using the larger $6\times6$ unit cell with projection. This is a manifestation of the hysteresis behavior.

\par
\begin{figure}[t!]
\includegraphics[width=0.45\textwidth]{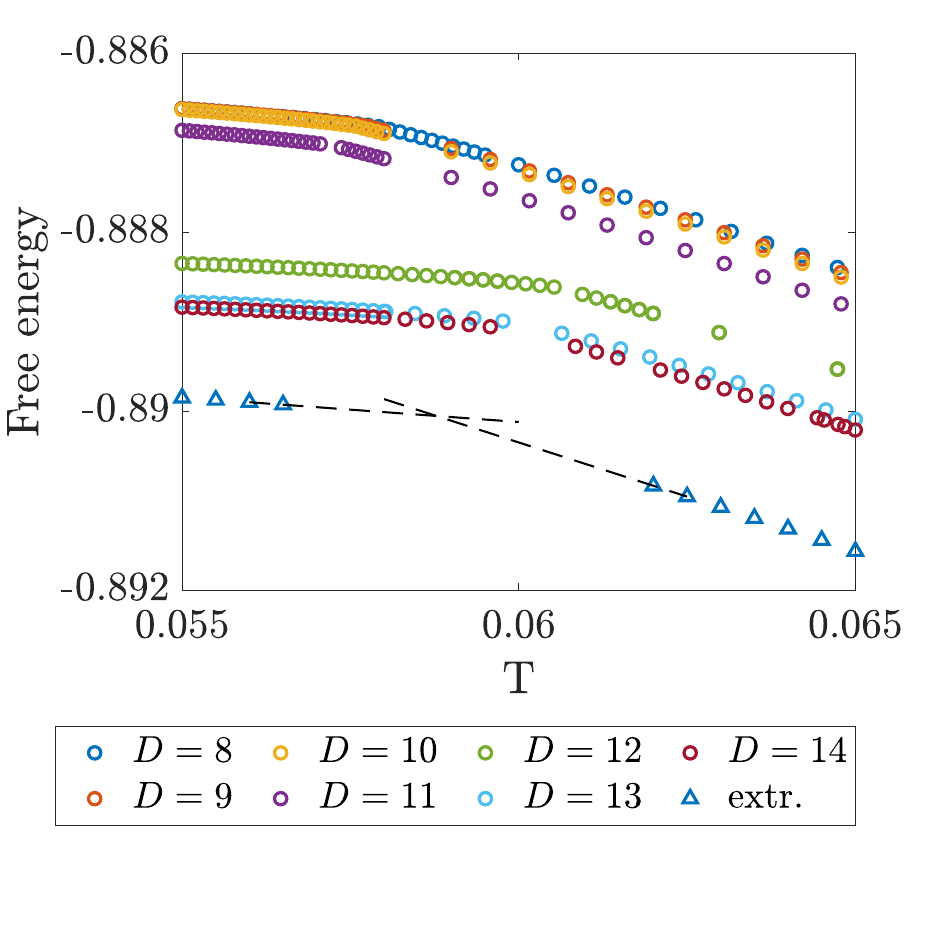}
\caption{Free energy with respect to the temperature for different bond dimensions $D$ with the extrapolated free energy (denoted by triangles). Using a linear extrapolation we find $T_c \simeq 0.0588$.}
\label{fig:free_ener_withextrap}
\end{figure}

 Next, we study the transition for different bond dimensions, summarized in Fig. \ref{fig:free_ener_withextrap}. For every considered $D$, we still observe a distinct kink in the free energy in agreement with a first-order transition. We note that by considering larger bond dimensions $D'>D$, the free energy decreases, $i.e.$ $f_D' < f_D$ conforming to the idea that a larger bond dimension will represent the Gibbs state more accurately. Although the free energy decreases systematically when considering a larger bond dimension, the transition temperature obtained by locating the kink does not behave monotonically with $D$, such that we cannot extrapolate directly $T_c(D)$. In particular we found $\min_{D}T_c(D) = 0.0572$ and $\max_D T_c(D) = 0.0605$ reached for $D = 11$ and $D = 12$ respectively. 

\par
\begin{figure}[t!]
\includegraphics[width=0.45\textwidth]{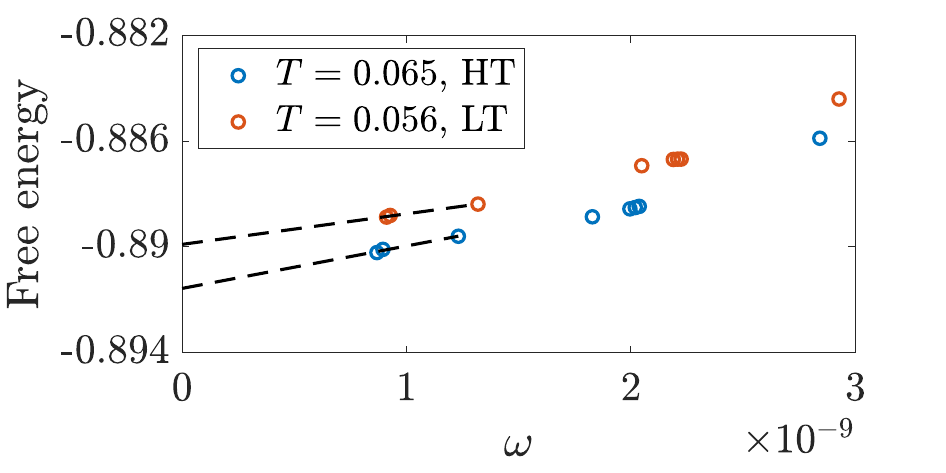}
\caption{Extrapolation of the free energy with respect to the truncation error for different bond dimensions ranging from $D = 7$ to $D = 14$. The free energy plotted with respect to $\omega$ is smoother than if plotted versus $1/D$ (not shown here).}
\label{fig:extrap_omega2}
\end{figure}

 In order to extract the critical temperature in the infinite $D$ limit, we first extrapolate the free energy in the ordered and disordered phases close to the transition and then locate the kink assuming the extrapolated free energy to behave linearly near the critical temperature. The finite$-D$ free energies are extrapolated with respect to the truncation error $\omega$ at the last iteration as done in Ref. ~\cite{Corboz2016}. Within the SU approach it can be computed as 
\begin{align}
\omega = \sum_{i \in \text{gates}} \omega_{i}
\end{align}
where the sum is taken over all gates applied during the last time step and 
\begin{align}
\omega_i = \sum_{k>D} s_k^2 / \sum_{k} s_k^2
\end{align}
with $s_k$ denoting the singular values associated to the singular value decomposition following the application of the gate $i$. In practice, a spline interpolation with respect to the temperature is first performed on the weights and free energies which are then extrapolated for a given set of temperatures. The extrapolations for two different temperatures in the ordered and disordered phases are shown in Fig. \ref{fig:extrap_omega2}. We can see that the results are reasonably smooth with respect to $\omega$ and assuming that we are in a linear regime we extrapolate the free energy by performing a linear regression on the highest three values of $D$ at our disposal. Finally we identify the first-order transition temperature around $T_c \simeq 0.0588$ (Fig. \ref{fig:free_ener_withextrap}), which is within the range of critical temperatures $T_c(D)$ obtained with finite bond dimension $D$.

\subsection{With pinning field $h_s > 0$}

\begin{figure}[t!]
\includegraphics[width=0.45\textwidth]{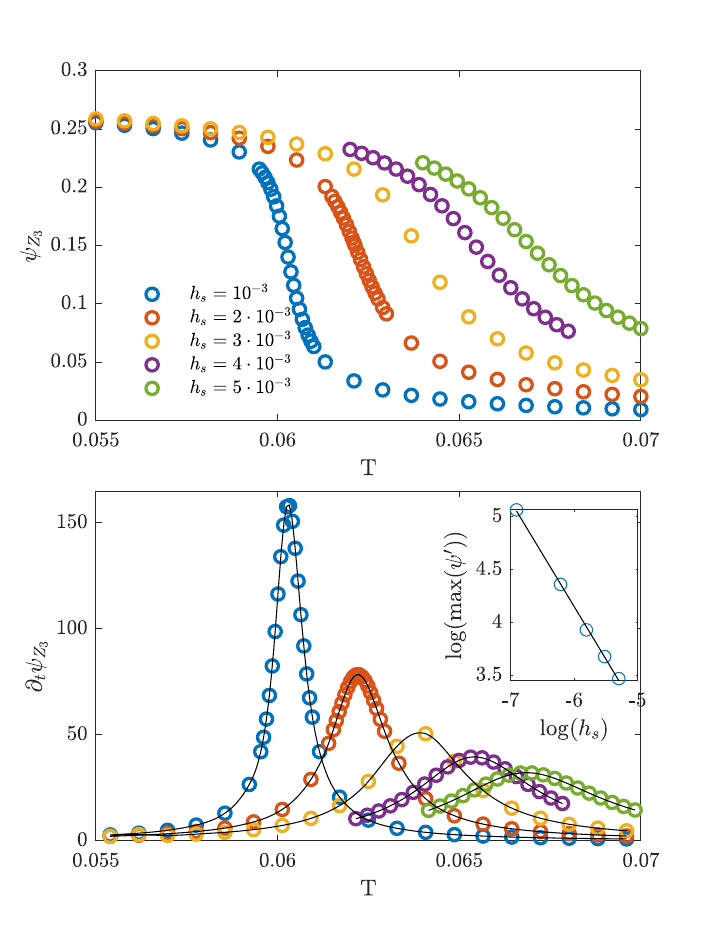}
\caption{Top panel: order parameter for various pinning fields. Bottom panel : The temperature derivative of the order parameter $\psi'$. Inset: log-log plot of the maximum of $\psi ' $ with respect to the pinning field. The computations were performed for $D = 8$ and boundary bond dimension $\chi = 40$.}
\label{fig:scalingD8_vert}
\end{figure}

In this section, the imaginary time evolution is performed on the $6\times 6$ unit cell. We first discuss in details the results obtained for $D = 8$. The order parameter and its temperature derivative $\psi' = \partial_t \psi_{Z_3}$ are shown in Fig.~\ref{fig:scalingD8_vert}. In the upper panel, we observe, as expected, that the order parameter smooths out while increasing the pinning field. On the lower panel we observe that the maximum of $\psi'$ scales linearly with the inverse of the pinning field $h_s^{-1}$, in agreement with a first-order transition. More precisely, by fitting $\log(\max \psi')$ linearly with respect to $\log(h_s)$ we found a slope of $0.996$. The linear fit is shown in the inset of Fig. \ref{fig:scalingD8_vert} where we have computed $\max \psi'$ by considering the maximum of the spline interpolation shown with black lines on the lower panel.

\begin{figure}[t!]
\includegraphics[width=0.45\textwidth]{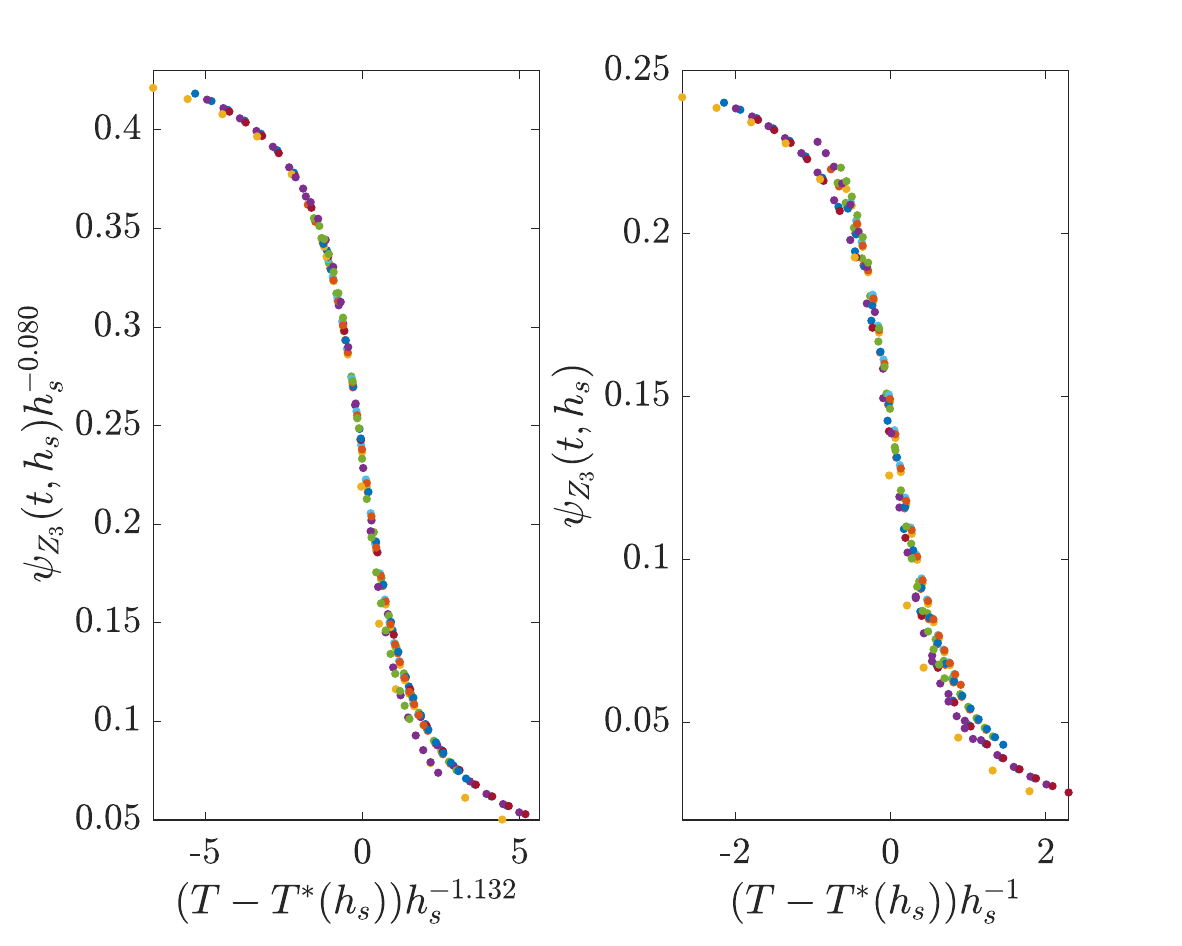}
\caption{The left and right panels show two collapses for the order parameter $\psi_{Z_3}$ performed with different critical exponents. In particular, the order parameter on the left panel is collapsed with critical exponents $\beta = 0.071$ and $\nu =  0.477$, that have been found to lead to an optimal collapse, while the order parameter on the right panel is collapsed with critical exponents $\beta = 0$ and $\nu =  1/2$ which are predicted for first-order transitions.}
\label{fig:scalingAllD}
\end{figure}

Next, we try to verify the scaling (\ref{scaling}) by collapsing the order parameter for various bond dimensions and pinning field values. We show the results in Fig. \ref{fig:scalingAllD} where we have collapsed the order parameter for $D$ ranging from $D=9$ to $D = 12$ and $h_s \in \{ 10^{-3}, 2\cdot 10^{-3}, 3\cdot 10^{-3}\}$. In the right panel, we show the collapse done with critical exponents predicted by a first-order transition $\beta = 0$ and $\nu = 1/2$ while on the left we show what we found to be the best collapse with critical exponents $\beta = 0.071$ and $\nu = 0.477$. Those exponents are obtained in two steps. First, for every bond dimension $D$, we fit the best exponents $\nu_D(h_s)$ and $\beta_D(h_s)$ collapsing the magnetization for two consecutive fields. Then, $\nu$ and $\beta$ are obtained by taking the average over $\nu_D(h_s)$ and $\beta_D(h_s)$ respectively.
The discrepancy between the theoretical exponents and the one best collapsing the order parameter is small enough such that the scaling is consistent with a first-order transition. We discuss the possible causes for this discrepancy in the next paragraphs.

\begin{figure}[t!]
\includegraphics[width=0.45\textwidth]{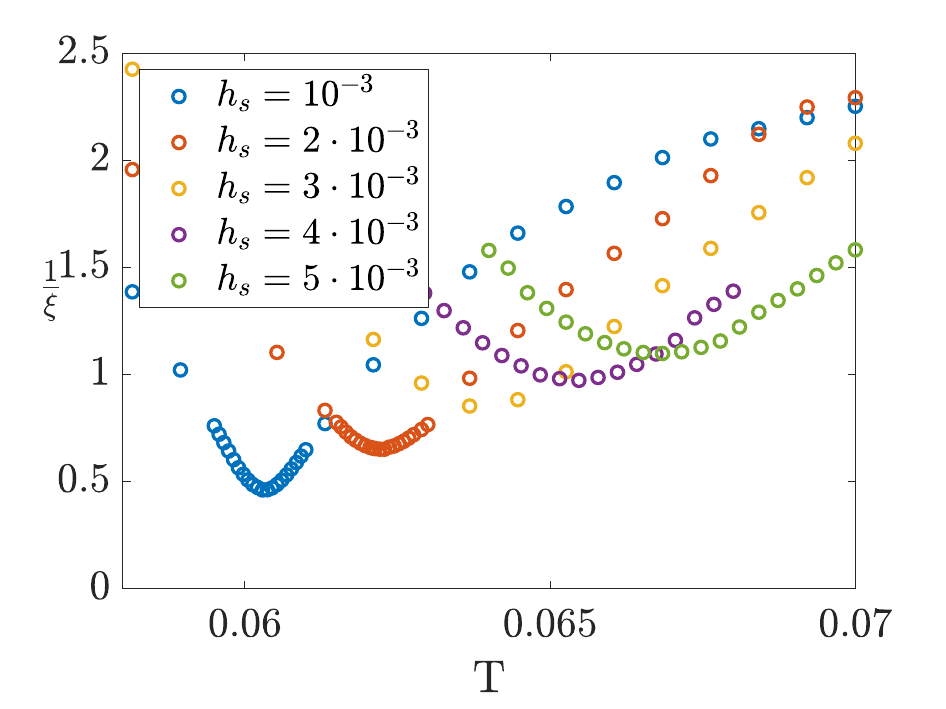}
\caption{Correlation length for various pinning fields for $D = 8$ with respect to the temperature. We observe that the correlation length peaks at the crossover temperature and that its maximum becomes larger when $h_s$ decreases.}
\label{fig:Xi}
\end{figure}

We now look at the correlation length for $D = 8$ and various pinning fields and show the results in Fig. \ref{fig:Xi}. We observe that the correlation length peaks at the crossover temperature $T^*(h_s)$ and that it increases when $h_s\rightarrow 0$. It is worth mentioning that if the transition was continuous, one would also numerically observe the correlation length to reach its maximum at the critical temperature and we are unable, based solely on the correlation length data, to rule out its divergence. However, this scenario is inconsistent with the kink observed in the free energy and allows us to conclude for the transition to be weakly first-order, similarly to the six-state Potts model. 

For two dimensional finite-size systems, the order parameter scales like $m(t, L) =  m(L^{2}t)$ near a first-order transition. As a result, the finite size of the system smooths out the order parameter with a slope at the transition proportional to $L^{2}$ and the pinning field can be interpreted as an effective length with $h_s^{-1} = L^2$. Moreover, the finite-size scaling is only valid for system sizes larger then the correlation length at the transition, $i.e.$ $L> \xi_c$ and it is notoriously difficult to study weakly first-order transition for systems with very large correlation length at the critical temperature. Hence, for the finite-pinning field scaling to be valid one would need $h_s^{-2} > \xi_c$. If we assume the correlation length at the transition to be of the order of the six-state Potts model $\xi_c = 159$ we would need a field of the order of $h_s \sim O(10^{-5})$ for the finite-pinning field scaling to be valid, two orders of magnitudes below the range studied here. To remedy the discrepancy in the critical exponents, we then would just need to decrease the pinning field and look at how the scaling behaves. However, as we decrease the pinning field, the correlation length increases and we expect the finite-$D$ effect to become larger and add extra corrections to the scaling. It is worth noting that already for $h_s \sim O(10^{-3})$ the order parameter has not converged in $D$ and the finite-$D$ effect are non negligible. Thus, the discrepancy between the theoretical and best fitted exponents is probably due to a mix of finite-$D$ effect and the pinning field being too large for the scaling to be entirely valid. 

\begin{figure}[t!]
\includegraphics[width=0.45\textwidth]{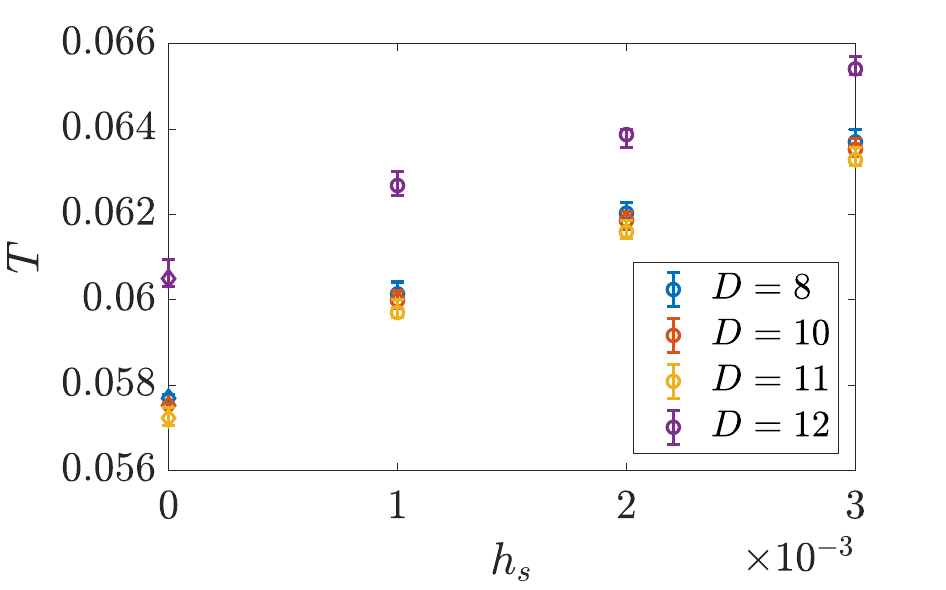}
\caption{Crossover temperatures for different pinning fields and bond dimensions. The critical temperature $T_c(D)$ obtained by looking at the free energy is denoted by the diamond markers as opposed to the circle markers which denote $T^*_D(h_s)$.}
\label{fig:scalingTc}
\end{figure}

Finally we can look at how the crossover temperature approaches the critical temperature for different bond dimensions $D$. The results are summarised in Fig. \Ref{fig:scalingTc}. $T^*_D(h_s)$ is estimated by determining at which temperature $\partial_t \psi_{Z_3}$ reaches its maximum for some bond dimension, while $T_c(D)$ is obtained by linearly extrapolating the free energy near the transition. We observe that the finite field results are consistent with the one obtained without pinning field even though they are measured by different methods. We recall that $T^*(h_s)$ should approach $T_c$ linearly with respect to the pinning field. However, a linear regression on the smallest field available for $D = 9,10,11$ slightly overestimates the critical temperature. This again indicates that we are not yet in the regime where the scaling is entirely valid. For $D = 12$ it is harder to conclude as the error bars are larger but the results are consistent with $T^*_D(h_s)$ approaching $T_c(D)$ linearly.

\subsection{Phase diagram}

\begin{figure}[t!]
\includegraphics[width=0.45\textwidth]{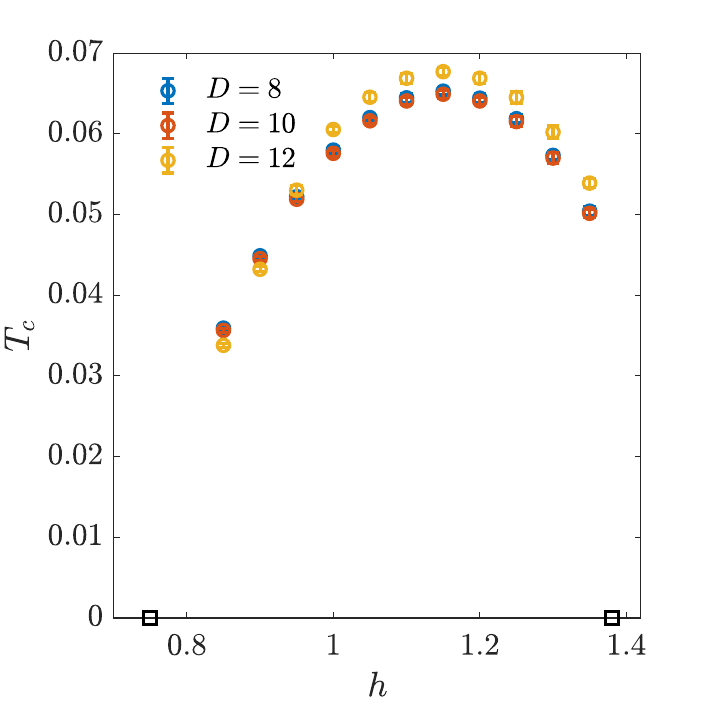}
\caption{Finite temperature phase diagram for the melting of the 1/3 plateau. The black squares indicate the extent of the plateau at zero temperature.}
\label{fig:dome_shape}
\end{figure}

Finally, we map out the melting of the 1/3 plateau as a function of $h$ for different bond dimensions $D$ and show the results in Fig. \ref{fig:dome_shape}. We observe the ordered phase to follow a dome shape. Moreover, the $D = 8,10$ and $D=12$ lines cross around $h \simeq 0.92$ indicating that the finite-$D$ effects change along the transition line and vary with the field. The extent of the 1/3 plateau at zero temperature found in Ref. \cite{Mila2013b} is indicated with black squares on the plot for comparison. 

\section{Discussion and Summary}

We have investigated the nature of the melting of the 1/3 plateau in the Shastry-Sutherland model and how the broken $\mathbb{Z}_2$ and $\mathbb{Z}_3$ symmetries are restored at high temperature. In particular we found both symmetries to be restored at a unique temperature at which the free energy exhibits a kink for different bond dimensions. We further found the correlation length to peak at the critical temperature, thus showing the transition to be weakly first-order similarly to the six-state Potts model. Also, we investigated the scaling induced by a pinning field and found the order parameter to collapse with critical exponents that agree with a first-order transition. Furthermore, by extrapolating the finite-$D$ free energy with respect to the truncation error the transition is found to occur around $T_c \simeq 0.0588$. Using the value $J = 84$K measured in Ref. \cite{honecker2019} we obtain a critical temperature of about $T_c = 4.8$K well above the temperature of 2K at which the 1/3 plateau was observed experimentally. Finally, we have provided a systematic methodology to investigate first-order transitions in 2D systems with iPEPS at finite temperature.\\

{\it Acknowledgments.} SN thanks Olivier Gauthé for useful discussions.
This work has been supported by the Swiss National Science Foundation Grant No. 212082.
The calculations have been performed using the facilities of the Scientific IT and Application Support Center of EPFL. This project has received funding from the European Research Council (ERC) under the European
Union’s Horizon 2020 research and innovation programme
(grant agreement No. 101001604).

\appendix

\section{Fast Full Update}
\label{section_ffu}

In this section we use the FFU algorithm to perform the imaginary time evolution and compare the results with the one obtained by the SU scheme. Instead of truncating a bond by performing a singular value decomposition, the FFU looks for the new tensors that best approximate the update by taking into account the environment computed via CTMRG. This in turn improves the accuracy of the imaginary time evolution but becomes computationally much more expensive. To mitigate this drawback, it is possible to combine the two approaches and use a mix of SU and FFU \cite{Jacek2021} where the SU is used to perform the evolution up to an inverse temperature $\beta_{SU}$ and then, the FFU is used to perform the evolution for the remaining steps. For a complete discussion on the fast full update we refer the reader to Ref. \cite{Orus2015}. Here, we use a combination of the SU and FFU schemes with $\beta_{SU} = 10$ on the $6\times 6$ unit cell shown in Fig. \ref{fig:peps_unit_cell} and project the boundary condition as discussed in Section \ref{bc_sec}. Similarly to the SU, in order to compute the free energy with the FFU scheme, we need to keep track of the renormalisation factors.

\begin{figure}[t!]
\includegraphics[width=0.45\textwidth]{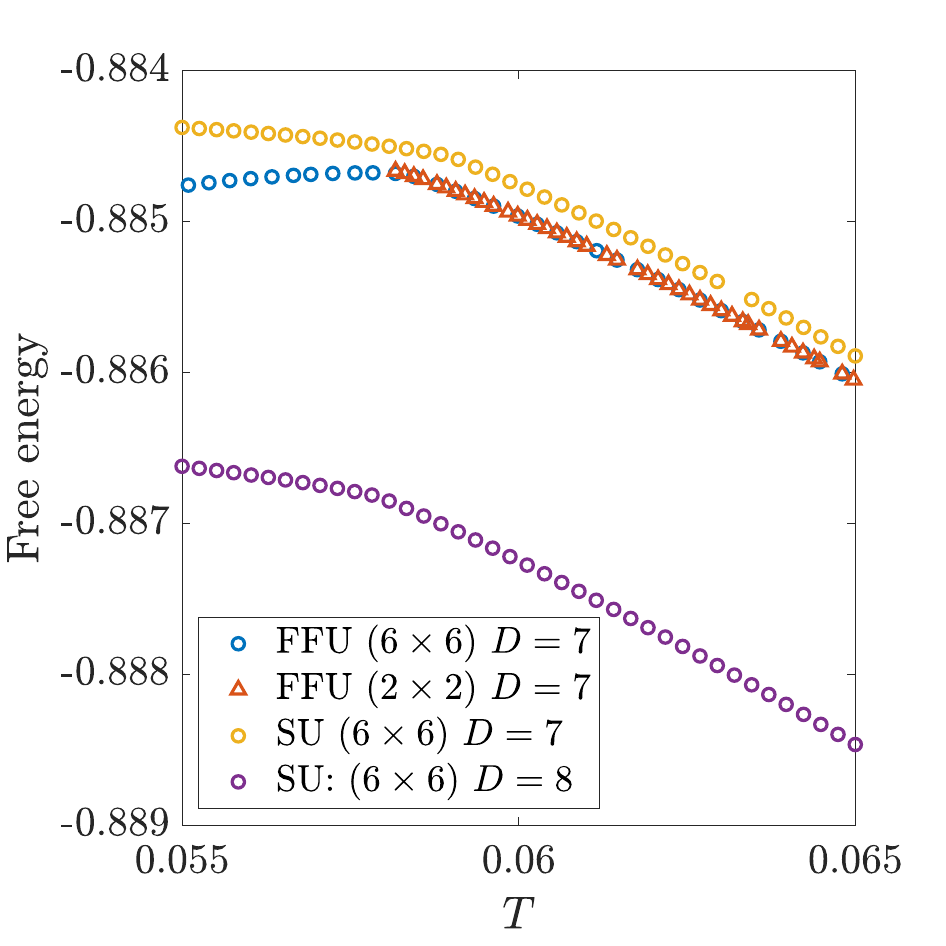}
\caption{Comparison between the SU and FFU schemes done with $\chi = 40$ and $\chi = 50$ respectively.}
\label{fig:FFU}
\end{figure}

The results are summarised in Fig. \ref{fig:FFU} where we used a time step $d\beta = 0.02$ for the FFU. We observe that for $D=7$, the free energies computed via both schemes exhibit a kink around the same temperature in agreement with a first-order transition. We further check that the results with the FFU performed using the $2\times 2$ unit cell and $6\times 6$ unit cell with projected boundary condition give the same free energy in the disordered phase. Moreover, the free energy computed with the FFU scheme is as expected smaller than the one computed with SU. However, by plotting the results obtained with SU for larger bond dimension, namely $D=8$, we note that the difference of free energy obtained from the SU and FFU is rather small compared to the finite-$D$ effect. The small free energy difference between the two schemes, which is probably due to the absence of a diverging correlation length in the system, justifies using the SU approach, which can be pushed to considerably larger bond dimensions than the FFU method.

\section{Classical Ising model on the Shastry-Sutherland lattice}

\begin{figure}[t!]
\includegraphics[width=0.45\textwidth]{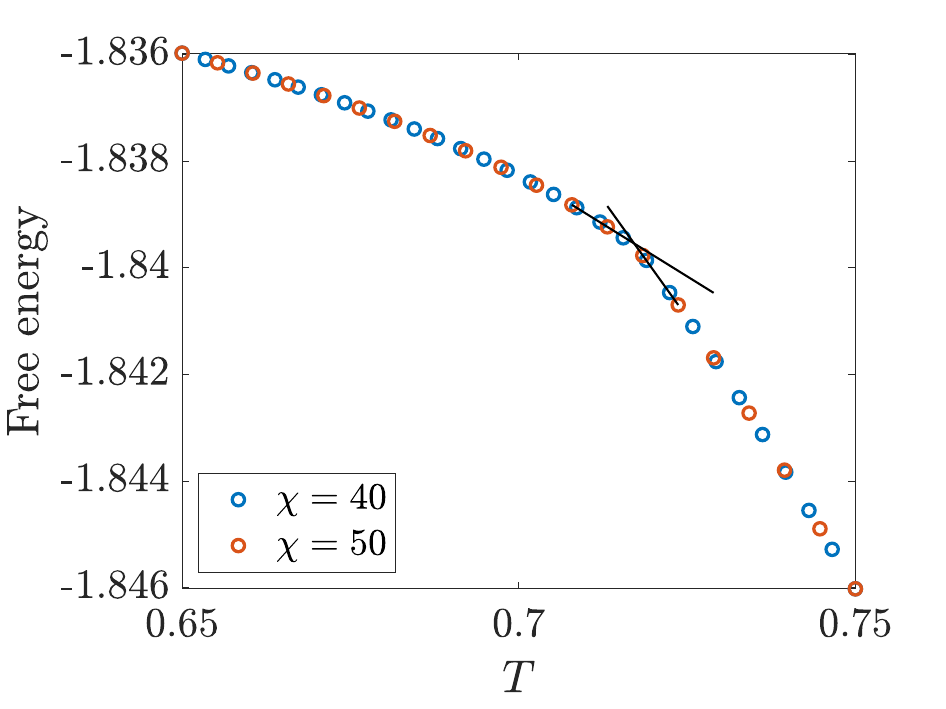}
\caption{Free energy with respect to the temperature for different bond dimension $\chi = 40$ and $\chi = 80$. The free energy exhibits a kink around $T_c \simeq 0.72$ in agreement with a first-order transition. The black lines are linear fits of the two closest points to the transition, in the ordered and disordered phase, for $\chi = 50$.}
\label{fig:free_ener_cl}
\end{figure}

In this section we discuss the classical Ising model on the Shastry-Sutherland lattice whose ground states also break the same rotational $\mathbb{Z}_2$ and translational $\mathbb{Z}_3$ symmetries. The Hamiltonian is given by :
\begin{align}
H = J' \sum_{\langle i,j \rangle} \sigma_i \sigma_j + J  \sum_{\langle \langle i,j \rangle\rangle} \sigma_i \sigma_j  - h \sum_i \sigma_i
\end{align}
with spins $\sigma \in {\pm 1}$. The model has already been studied with the Tensor Renormalisation Group (TRG) algorithm in Ref. \cite{Fong2009} where the structure of the ground state and the location of the transition is discussed. However, the nature of the melting is not. By mapping the partition function on a two-dimensional tensor network and performing the contraction with CTMRG we can study the nature of the transition. We map the partition function on the bipartite lattice shown in Fig. \ref{fig:cl_tensors} composed of tensors $v$ and $h$ given by:
\begin{figure}[t!]
\includegraphics[width=0.45\textwidth]{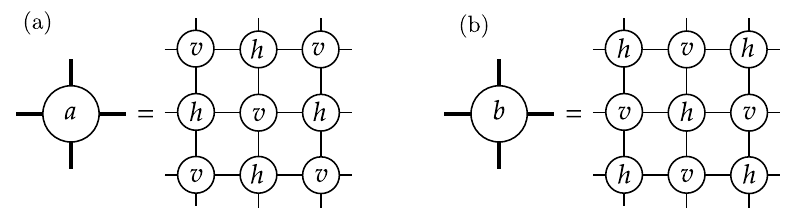}
\caption{Local tensors used to compute the partition function. The bold lines represent bond dimension 8 while the thin lines represent bond dimension 2. The local tensors $h$ and $v$ are given by Eqs. \ref{eq1} and \ref{eq2}.}
\label{fig:cl_tensors}
\end{figure}

\begin{align}
& (v)_{\sigma_1,\sigma_2,\sigma_3,\sigma_4}  \propto \text{exp} ( -K_1 (\sigma_1 \sigma_2 + \sigma_2 \sigma_3 + \sigma_3 \sigma_4 +\ldots  \nonumber \\
&  \quad   + \sigma_4 \sigma_1) - K_2 \sigma_1 \sigma_3 + \frac{H}{2} (\sigma_1 + \sigma_2 + \sigma_3 + \sigma_4)  ) \label{eq1} \\
& (h)_{\sigma_1,\sigma_2,\sigma_3,\sigma_4}  \propto \text{exp}  ( -K_1 (\sigma_1 \sigma_2 + \sigma_2 \sigma_3 + \sigma_3 \sigma_4 +\ldots  \nonumber \\
&  \quad   + \sigma_4 \sigma_1) - K_2 \sigma_2 \sigma_4 + \frac{H}{2} (\sigma_1 + \sigma_2 + \sigma_3 + \sigma_4)  ) 
\label{eq2}
\end{align}
with $K_1 = \beta J', K_2 = \beta J$ and $H = \beta h$. We focus on $J = J' = 1$ and $h = 3$ which is into the 1/3 plateau \cite{Dublenych2012,Fong2009} and show the results in Fig. \ref{fig:free_ener_cl}. The simulations were performed using fixed boundary conditions, favouring one out of the ground states. Similarly to the quantum case, we observe a single kink in the free energy around $T_c \simeq  0.718$, thus showing the transition to be first-order. Although not shown here, we have further checked that the $\mathbb{Z}_2$ and $\mathbb{Z}_3$ symmetries are restored at the same temperature. It is worth noting that Ref. \cite{Fong2009} has a different convention and uses spin one-half so that a factor four is needed to recover a comparable critical temperature. 

\bibliographystyle{apsrev4-1}
\bibliography{bibliography,comments}

\end{document}